\documentclass[12pt,a4paper]{article}
\usepackage{epsfig}
\usepackage{amsmath,amsfonts,amssymb}
\usepackage{t1enc}
\usepackage{cite}
\newcommand{\ptmiss}{p_T\!\!\!\!\!\!\!\!\! \not \,\,\,\,\,\,\,\,}

\begin{document}
\vspace*{-3cm}
\begin{flushright}
hep-ph/0404104 \\
April 2004
\end{flushright}

\begin{center}
\begin{Large}
{\bf CP violation in $\boldsymbol{\tilde \chi_1^0 \tilde \chi_2^0}$ production
at a linear collider}
\end{Large}

\vspace{0.5cm}
J. A. Aguilar--Saavedra \\[0.2cm] 
{\it Departamento de Física and GTFP, \\
  Instituto Superior Técnico, P-1049-001 Lisboa, Portugal} \\
\end{center}

\begin{abstract}
We discuss the observability of CP-violating asymmetries in the process
$e^+ e^- \to \tilde \chi_1^0 \tilde \chi_2^0 \to \tilde \chi_1^0 \tilde
\chi_1^0 \ell^+ \ell^-$, with $\ell = e,\mu$. We consider two examples of
supersymmetric scenarios: ({\em i\/}) with decays
$\tilde \chi_2^0 \to \tilde \ell_R^\pm \ell^\mp \to \tilde \chi_1^0 \ell^+
\ell^-$; ({\em ii\/}) with $\tilde \chi_2^0$ three-body decays.
The asymmetries can be of order 0.1 but they are partially
washed out by the large backgrounds from $W^+ W^-$ and slepton
pair production, being the observed asymmetries one order of magnitude
smaller. However, with appropriate kinematical cuts
they can be observed at an $e^+ e^-$ collider with a centre of mass
energy of 500 GeV and high luminosity.
\end{abstract}



\section{Introduction}
\label{sec:1}

There are several motivations to consider further CP violation
sources in addition to the CP-violating phase in the Cabibbo-Kobayashi-Maskawa
matrix. From
the experimental point of view, the Standard Model (SM) is unable to explain the
observed baryon asymmetry of the universe. From the theoretical side, most SM
extensions introduce new CP violation sources. The Minimal Supersymmetric
Standard Model (MSSM) \cite{susy2,susy3} contains several new phases which can
lead to observable
effects at high energy colliders. In the neutralino sector these are the phases
of the parameters $M_1$ and $\mu$,
$\phi_1$ and $\phi_\mu$ respectively. Large phases $\phi_1$ and/or
$\phi_\mu$ lead to supersymmetric contributions
to electric dipole moments (EDMs) far above present limits. However, these two
phases can be large without
necessarily yielding unacceptably large EDMs, if there are large
cancellations between the different contributions \cite{pr1,pr2,kane}.
One of the tasks to be accomplished at a future linear collider is to
explore the effects of these phases in phenomenology, in order to determine
if they vanish or not.

In this paper we study the process of $\tilde \chi_1^0 \tilde \chi_2^0$
production with subsequent leptonic decay of the second neutralino
\cite{choi0,bartl1,bartl2},
\begin{align}
e^+ e^- \to \tilde \chi_1^0 \tilde \chi_2^0 \to \tilde \chi_1^0 \tilde
\chi_1^0 \ell^+  \ell^- \,,
\label{ec:XX}
\end{align}
with $\ell = e,\mu$.\footnote{We do not consider $\tilde \chi_2^0 \to \tilde
\chi_1^0 \tau^+ \tau^-$, which is the dominant channel in some cases, because
each of the $\tau$ leptons decays producing one or two undetected neutrinos and
the reconstruction of the $\tau$ momenta is not possible. The study of a
CP-violating asymmetry involving the $\tau$ decay products requires a simulation
of the $\tau$ decay and is beyond the scope of this work.}
In this process, it is possible to have a CP asymmetry in the
triple product $\vec p_{e^+} \cdot ( \vec p_{\ell^-} \times \vec p_{\ell^+} )$
of order
0.1 for adequate choices of beam polarisations. This asymmetry is sensitive to
the phases of $M_1$ and $\mu$ and is due to of the influence
of $\tilde \chi_2^0$ polarisation in the $l^+$, $l^-$ angular
distributions (the expressions of the polarised matrix
elements for $\tilde \chi_1^0 \tilde \chi_2^0$ production and $\tilde \chi_2^0$
decay can be found in Refs.~\cite{gudrid1,gudrid2}).
However, its experimental detection
is jeopardized by the presence of huge backgrounds from the
production of $W^+ W^-$, selectron/smuon and, to a lesser extent, chargino
pairs,
\begin{align}
e^+ e^- & \to W^+ W^- \to \ell^+ \nu_\ell \, \ell^- \bar \nu_\ell \,, 
  \nonumber \\
e^+ e^- & \to \tilde \ell_{R,L}^+ \, \tilde \ell_{R,L}^- \to \ell^+ \tilde
  \chi_1^0 \, \ell^- \tilde \chi_1^0 \,, \nonumber \\
e^+ e^- & \to \tilde \chi_1^+ \tilde \chi_1^- \to \ell^+ \nu_\ell \tilde
  \chi_1^0 \, \ell^- \bar \nu_\ell \tilde \chi_1^0 \,,
\label{ec:XXb}
\end{align}
which give the same experimental signature of two oppositely charged leptons
$\ell^+ \ell^-$ plus missing energy and momentum. A realistic analysis taking
these backgrounds into account is compulsory in order to draw a conclusion
on the observability of this CP asymmetry.

We analyse two kinds of supersymmetry (SUSY) scenarios, depending on the
dominant channel contributing to the decay of the second neutralino:
({\em i\/}) scenarios with decays
$\tilde \chi_2^0 \to \tilde \ell_R^\pm \ell^\mp \to \tilde \chi_1^0 \ell^+
\ell^-$; ({\em ii\/}) scenarios where $\tilde \chi_2^0$ has three-body decays.
The case of $\tilde \chi_2^0 \to \tilde \ell_L^\pm \ell^\mp \to \tilde
\chi_1^0 \ell^+ \ell^-$ is similar to the decay
to $\tilde \ell_R^\pm \ell^\mp$ but involves a heavier neutralino spectrum,
for which the signal cross sections are smaller.
We do not study scenarios with decays $\tilde \chi_2^0 \to \tilde
\chi_1^0 Z \to \tilde \chi_1^0 \ell^+ \ell^-$ because the asymmetries are
rather small \cite{bartl2}, and turn out to be unobservable due to the large
backgrounds.  Instead of performing a scan over some region of the SUSY
parameter space (for such analysis see
Ref. \cite{bartl1}, where a study complementary to this one is
performed but
without including backgrounds nor the effect of ISR and beamstrahlung), we
concentrate on the detailed analysis of two specific
examples, to illustrate each of the two situations.
We consider $e^+ e^-$ annihilation at a centre of mass (CM) energy of 500 GeV,
as proposed for the first phase of TESLA.

This paper is organised as follows. In Section \ref{sec:2} we examine how
CP asymmetries can be defined in $\tilde \chi_1^0 \tilde \chi_2^0$
production, and fix the SUSY
scenarios to be discussed. In Section \ref{sec:3} we analyse in detail the
triple product CP asymmetry in two scenarios. The results
for other scenarios are also commented. In Section \ref{sec:4} we 
summarise our results and compare with other CP violation asymmetry which is
also sensitive to the phases $\phi_1$, $\phi_\mu$. In the Appendix we collect
some Lagrangian terms required for our calculations.

\section{CP asymmetries in $\boldsymbol{\tilde \chi_1^0 \tilde \chi_2^0}$
production and decay}
\label{sec:2}

\begin{figure}[htb]
\begin{center}
\begin{tabular}{ccccc}
\mbox{\epsfig{file=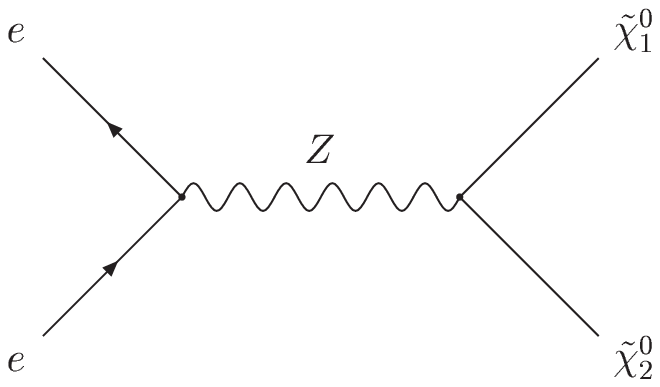,width=3.8cm,clip=}} & \hspace*{5mm} & 
\mbox{\epsfig{file=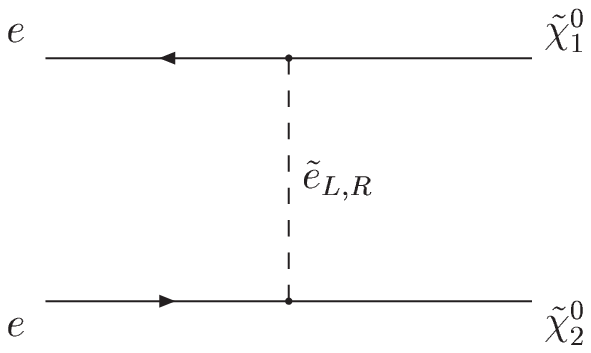,width=3.8cm,clip=}} & \hspace*{5mm} & 
\mbox{\epsfig{file=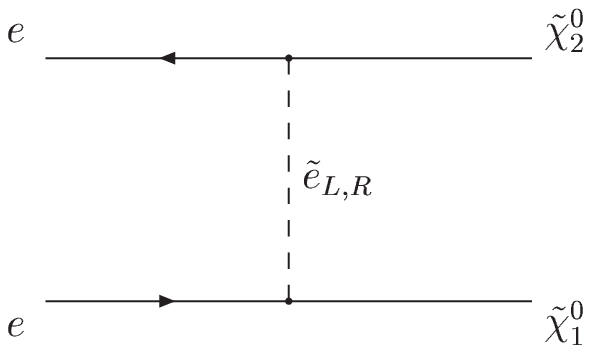,width=3.8cm,clip=}}
\end{tabular}
\caption{Feynman diagrams for $\tilde \chi_1^0 \tilde \chi_2^0$ production
in $e^+ e^-$ annihilation.}
\label{fig:XXprod}
\end{center}
\end{figure}

\begin{figure}[htb]
\begin{center}
\begin{tabular}{ccccc}
\mbox{\epsfig{file=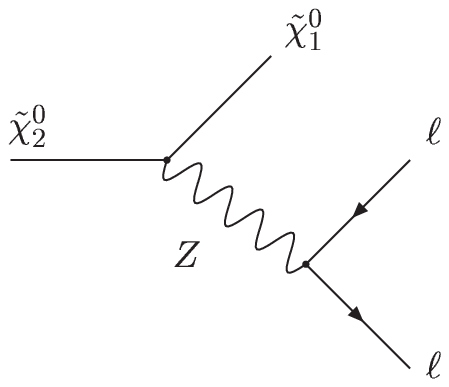,width=3cm,clip=}} & \hspace*{5mm} & 
\mbox{\epsfig{file=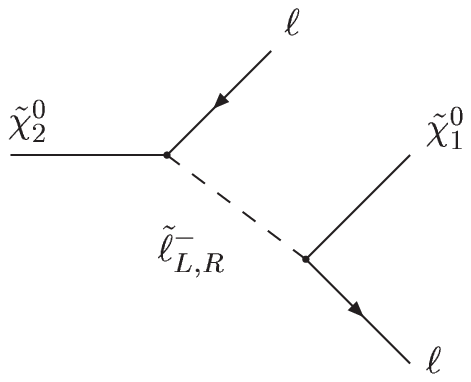,width=3cm,clip=}} & \hspace*{5mm} & 
\mbox{\epsfig{file=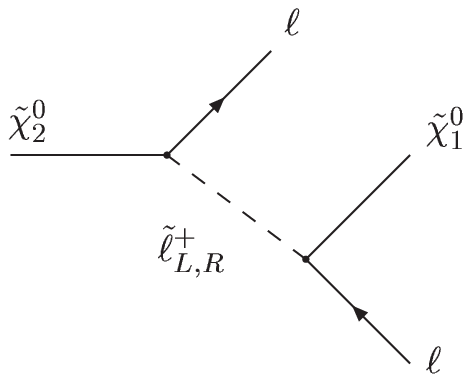,width=3cm,clip=}} \\
(a) & & (b) & & (c)
\end{tabular}
\caption{Feynman diagrams for the decay $\tilde \chi_2^0 \to \tilde \chi_1^0
\ell^+ \ell^-$.}
\label{fig:X2decay}
\end{center}
\end{figure}

The process of neutralino production in Eq.~(\ref{ec:XX}) takes place through
the diagrams depicted in Fig.~\ref{fig:XXprod}. The Lagrangian terms and
conventions used can be found in Ref.~\cite{npb} and the Appendix. 
From inspection of the diagrams and interactions involved we can notice that
only $e^+$ and $e^-$ with
opposite helicities give non-vanishing contributions to the amplitude
(neglecting selectron mixing), what constitutes a crucial point for the
construction of our CP asymmetries. The decay of $\tilde \chi_2^0$
is mediated by the diagrams in Fig. \ref{fig:X2decay} (production and decay
diagrams are shown separately only for clarity, in our computations we calculate
the complete matrix elements for the resonant process).
We have omitted the diagrams with neutral scalars, which are
proportional to $m_\ell$ and thus irrelevant for $\ell=e,\mu$.

\noindent

\begin{figure}[htb]
\begin{center}
\epsfig{file=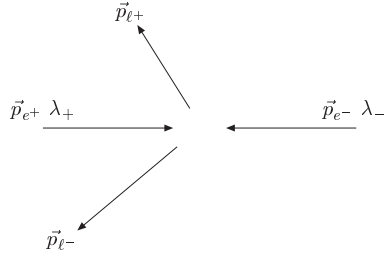,width=5cm,clip=}
\caption{Schematic picture of the process $e^+ e^- \to \tilde \chi_1^0
\tilde \chi_2^0 \to \tilde \chi_1^0 \tilde \chi_1^0 \ell^+ \ell^-$ in the CM
frame.}
\label{fig:xxprod}
\end{center}
\end{figure}

In the CM system the process  of $\tilde \chi_1^0 \tilde \chi_2^0$ production
looks as depicted in Fig.~\ref{fig:xxprod}, with the
3-momenta in obvious notation and $\lambda_- = \pm 1$, $\lambda_+ = -\lambda_-$
the electron and positron helicities, respectively.
The momenta of the two final state neutralinos
cannot be reconstructed because of the insufficient number of kinematical
constraints available for this process, thus we do not use them in our analysis.
Under CP, the momenta and helicities transform as
\begin{align}
& \vec p_{e^+} \to - \vec p_{e^-} = \vec p_{e^+} \,, \quad
\vec p_{e^-} \to - \vec p_{e^+} = \vec p_{e^-} \,, \quad
\vec p_{\ell^+} \to - \vec p_{\ell^-} \,, \quad
\vec p_{\ell^-} \to - \vec p_{\ell^+} \, \nonumber \\
& \lambda_+ \to -\lambda_- = \lambda_+ \,, \quad
\lambda_- \to -\lambda_+ = \lambda_-
\label{ec:trans}
\end{align}
(remember that in the CM frame $\vec p_{e^+} = - \vec p_{e^-}$). The
initial state is CP-symmetric independently of the possible beam polarisations,
owing to the fact that $\lambda_+ = - \lambda_-$. Therefore, the quantities
\begin{eqnarray}
Q_1 & = & \vec p_{e^+} \cdot \left( \vec p_{\ell^-} \times \vec p_{\ell^+}
\right) \,, \nonumber \\
Q_2 & = & \vec p_{e^+} \cdot \left( \vec p_{\ell^-} + \vec p_{\ell^+}
\right)
\end{eqnarray}
are CP-odd (other higher-order CP-odd quantities may also be built
using the vectors in Eqs.~(\ref{ec:trans})).
For $Q_{1,2}$ we define the asymmetries
\begin{equation}
A_i \equiv \frac{N(Q_i > 0) - N(Q_i < 0)}{N(Q_i > 0) + N(Q_i < 0)} \,,
\label{ec:asim}
\end{equation}
where $N$ denotes the number of events. These asymmetries must vanish if
CP is conserved, and are genuine signals of CP violation. Since $Q_2$ is
even under naive time reversal T, in order to have a nonvanishing asymmetry
$A_2$ the presence of CP-conserving phases in the amplitude is needed. In the
process under consideration, and neglecting the small phases originated from
particle widths, a nonzero $A_2$ arises from the interference of a
dominant tree-level and a subleading loop diagram mediating the decay.
Thus, $A_2$ is expected to be very small. On the other hand
$Q_1$ is T-odd, and relatively large asymmetries $A_1$ are possible already at
the tree level. We focus our analysis on the asymmetry $A_1$.
We note that particle-antiparticle identification is necessary in order to build
a triple product CP asymmetry in this process. In hadronic decays
$\tilde \chi_2^0 \to \tilde \chi_1^0 q \bar q$ one could try to build an
analogous asymmetry distinguishing the
two quark jets by their energy. However, a triple product such as
$\vec p_{e^+} \cdot \left( \vec p_{q_1} \times \vec p_{q_2} \right)$ is CP-even,
and at least three untagged jets in the final state are required to construct
a CP-odd triple product.

In the presence of a ``symmetric'' background the observed asymmetry 
$A_1^\mathrm{eff}$ is smaller than $A_1$
because the background does not contribute to the numerator of
Eq.~(\ref{ec:asim}) but contributes to the denominator. If we define the ratio
\begin{equation}
r = \frac{S}{S+B} \,,
\label{ec:r}
\end{equation}
$S$ and $B$ denoting the number of signal and background events, respectively,
the effective asymmetry and its statistical error are
\begin{equation}
A_1^\mathrm{eff} = r A_1 \,, \quad \Delta A_1^\mathrm{eff} \simeq \sqrt r
\Delta A_1 \,,
\label{ec:eff}
\end{equation}
where the statistical error of the signal alone is
$\Delta A_1 = \sqrt{(1-A_1^2)/S}$. The second relation in Eq.~(\ref{ec:eff})
holds to a very good accuracy for the values of $A_1$
found in this work. Then, with the presence of a background which
does not have a CP asymmetry, the statistical significance $A_1/\Delta A_1$
decreases by a factor $\sqrt r$.

Our first supersymmetric scenario is very similar to the scenario SPS1a
in Ref.~\cite{sps}. The parameters relevant for our analysis are collected in
Table~\ref{tab:sc}. They approximately correspond to $m_{1/2} = 250$ GeV,
$m_{\tilde E} = m_{\tilde L} = m_{H_i} = 100$ GeV, $A_E = -100$ GeV at the
unification scale, and $\tan \beta = 10$. In this scenario, the
diagrams dominating $\tilde \chi_2^0 \to \tilde \chi_1^0 \ell^+ \ell^-$ are
those with decay to on-shell sleptons $\tilde \ell_R^\pm$ in
Fig.~\ref{fig:X2decay}c, \ref{fig:X2decay}b. 
For a heavier slepton spectrum (and the same $\tilde \chi_2^0$ mass) two-body
decays are not kinematically allowed, and $\tilde \chi_2^0$ has three-body
decays. This corresponds to our second scenario, with
$m_{\tilde E} = m_{\tilde L} = m_{H_i} = 200$ GeV,  $A_E = -200$ GeV. 
In both scenarios $\tilde \chi_1^0$ and $\tilde \chi_2^0$ are gaugino-like and
$\tilde \chi_1^0$ is mainly a bino. We also comment on the the situation when 
there is more mixing in the neutralino sector, so that $\tilde \chi_1^0$ has a
sizeable wino component.

\begin{table}[htb]
\begin{center}
\begin{tabular}{ccccc}
Parameter & ~ & Scenario 1 & ~ & Scenario 2 \\
\hline
$M_1$ & & 101.8 $e^{i \phi_1}$ & & 102.0 $e^{i \phi_1}$ \\
$M_2$ & & 191.8 & & 192.0 \\
$\mu$ & & 358.5 $e^{i \phi_\mu}$ & & 377.5 $e^{i \phi_\mu}$ \\
$\tan \beta$ & & 10 & & 10 \\
$m_{\tilde e_R},m_{\tilde \mu_R}$ & & 142.5 & & 224.0 \\
$m_{\tilde e_L},m_{\tilde \mu_L}$ & & 200.7 & & 264.5 \\
$m_{\tilde \chi_1^0}$ & & 98.7 & & 99.1 \\
$m_{\tilde \chi_2^0}$ & & 176.4 & & 178.1
\end{tabular}
\caption{Low-energy parameters (at the scale $M_Z$) for the two SUSY scenarios
discussed.
The dimensionful parameters are in GeV. The neutralino masses correspond to
$\phi_1=0$, $\phi_\mu=0$, and differ less than $\pm 3$ GeV for other phases.
\label{tab:sc}}
\end{center}
\end{table}

We have checked that in both scenarios it is possible to have the electron EDM
$d_e$ below the present experimental limit
$d_e^\mathrm{\,exp} = (0.079 \pm 0.074) \times 10^{-26} ~ e$ cm
\cite{pdb}.\footnote{The neutron and Mercury atom EDMs do not impose a
constraint on our analysis, since the phase of the gluino mass $M_3$ is not
involved, and additionally because the experimental limits on these quantities
can be satisfied if the quark spectrum is heavy enough.}
Using the expressions for the electron EDM in Ref.~\cite{arnowitt}, we find that
for each value of $\phi_1$ between 0 and $2\pi$ it is possible to find a narrow
interval for $\phi_\mu$ (which can be chosen such that $|\phi_\mu| \leq
0.08$ in scenario 1 and $|\phi_\mu| \leq 0.12$ in scenario 2) in which
the chargino and neutralino contributions to $d_e$ cancel, resulting in a
value compatible with experiment. In our numerical calculations we let
$\phi_1$ vary freely and set $\phi_\mu = 0$, bearing in mind that their values
are strongly correlated but our CP asymmetries and cross sections
are insensitive to the small variation of $\phi_\mu$ in the ranges required for
the cancellations of the EDMs ($|\phi_\mu| \leq 0.08$, $|\phi_\mu| \leq 0.12$).

\section{Results}
\label{sec:3}

We calculate the matrix elements for the resonant processes in
Eqs.~(\ref{ec:XX},\ref{ec:XXb})
using {\tt HELAS} \cite{helas}, so as to include all spin correlations and
finite width effects. We assume a CM energy of 500 GeV
and an integrated luminosity of 345 fb$^{-1}$ per year
\cite{lum}. In our calculation we take into account the
effects of initial state radiation (ISR) \cite{isr} and beamstrahlung
\cite{peskin,BS2}, using for the latter the
design parameters $\Upsilon = 0.05$, $N = 1.56$ \cite{lum}.\footnote{The actual
expressions for ISR and beamstrahlung  used in our calculation can be found in
Ref.~\cite{npb}.}
We also include a beam energy spread of 1\%.
In order to simulate the calorimeter and tracking resolution, we perform a
Gaussian smearing
of the energies of electrons and muons using the 
specifications in the TESLA Technical Design Report \cite{tesla2}
\begin{equation}
\frac{\Delta E^e}{E^e} = \frac{10\%}{\sqrt{E^e}} \oplus 1 \% \;, \quad
\frac{\Delta E^\mu}{E^\mu} = 0.02 \% \, E^\mu \;,
\end{equation}
where the energies are in GeV and the two terms are added in quadrature. We
apply ``detector'' cuts on transverse momenta, $p_T \geq 10$ GeV, and
pseudorapidities $|\eta| \leq 2.5$, the latter corresponding to polar angles
$10^\circ \leq \theta \leq 170^\circ$. We also reject events in which the
leptons are not isolated, requiring a ``lego-plot'' separation
$\Delta R = \sqrt{\Delta \eta^2+\Delta \phi^2} \geq 0.4$.
We do not require specific trigger conditions, and we assume that the
presence of charged leptons with high transverse momentum will suffice.
For the Monte Carlo integration in 6-body phase space we use
{\tt RAMBO} \cite{rambo}.

\subsection{Scenarios 1 and 2}

\begin{figure}[htb]
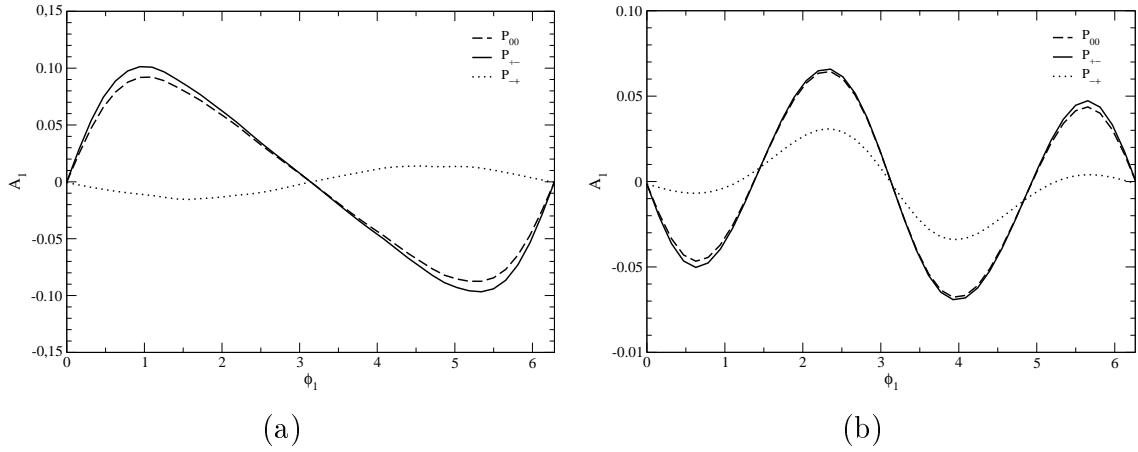

\begin{center}
\begin{tabular}{cc}
\epsfig{file=Figs/asimXX-2R.eps,width=7.3cm,clip=} 
& \epsfig{file=Figs/asimXX-3.eps,width=7.3cm,clip=} \\
(a) & (b)
\end{tabular}
\caption{Dependence of the asymmetry $A_1$ on the phase $\phi_1$ in scenario 1
(a) and scenario 2 (b). Backgrounds are not included. We consider unpolarised
beams ($P_{00}$), $P_{e^+}=0.6$, $P_{e^-}=-0.8$ ($P_{+-}$) and
$P_{e^+}=-0.6$, $P_{e^-}=0.8$ ($P_{-+}$).}
\label{fig:aXX}
\end{center}
\end{figure}

\begin{figure}[htb]
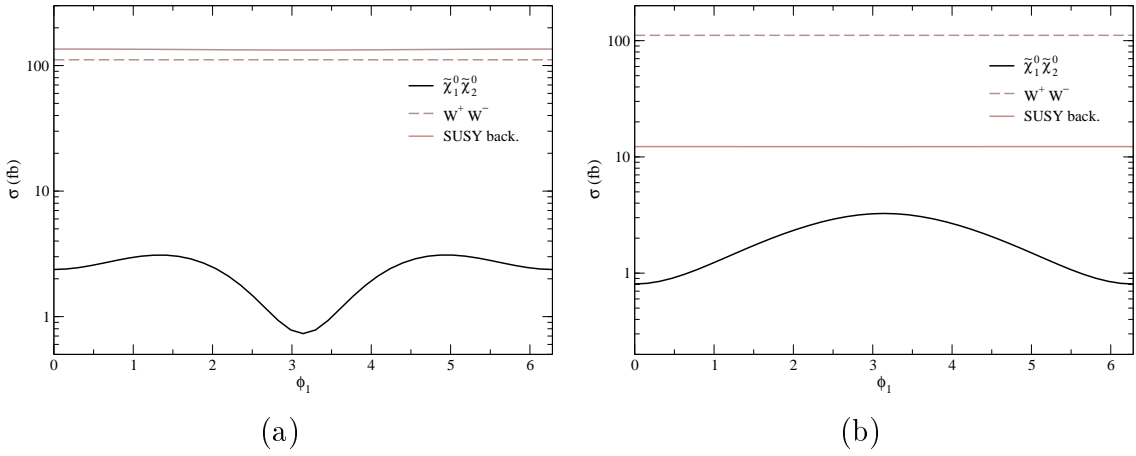

\begin{center}
\begin{tabular}{cc}
\epsfig{file=Figs/cross2R00.eps,width=7.3cm,clip=} 
& \epsfig{file=Figs/cross300.eps,width=7.3cm,clip=} \\
(a) & (b)
\end{tabular}
\caption{Dependence of the signal and background cross sections on the phase
$\phi_1$ in scenario 1 (a) and scenario 2 (b), for unpolarised beams.}
\label{fig:cr00}
\end{center}
\end{figure}

\begin{figure}[htb]
\begin{center}
\begin{tabular}{cc}
\epsfig{file=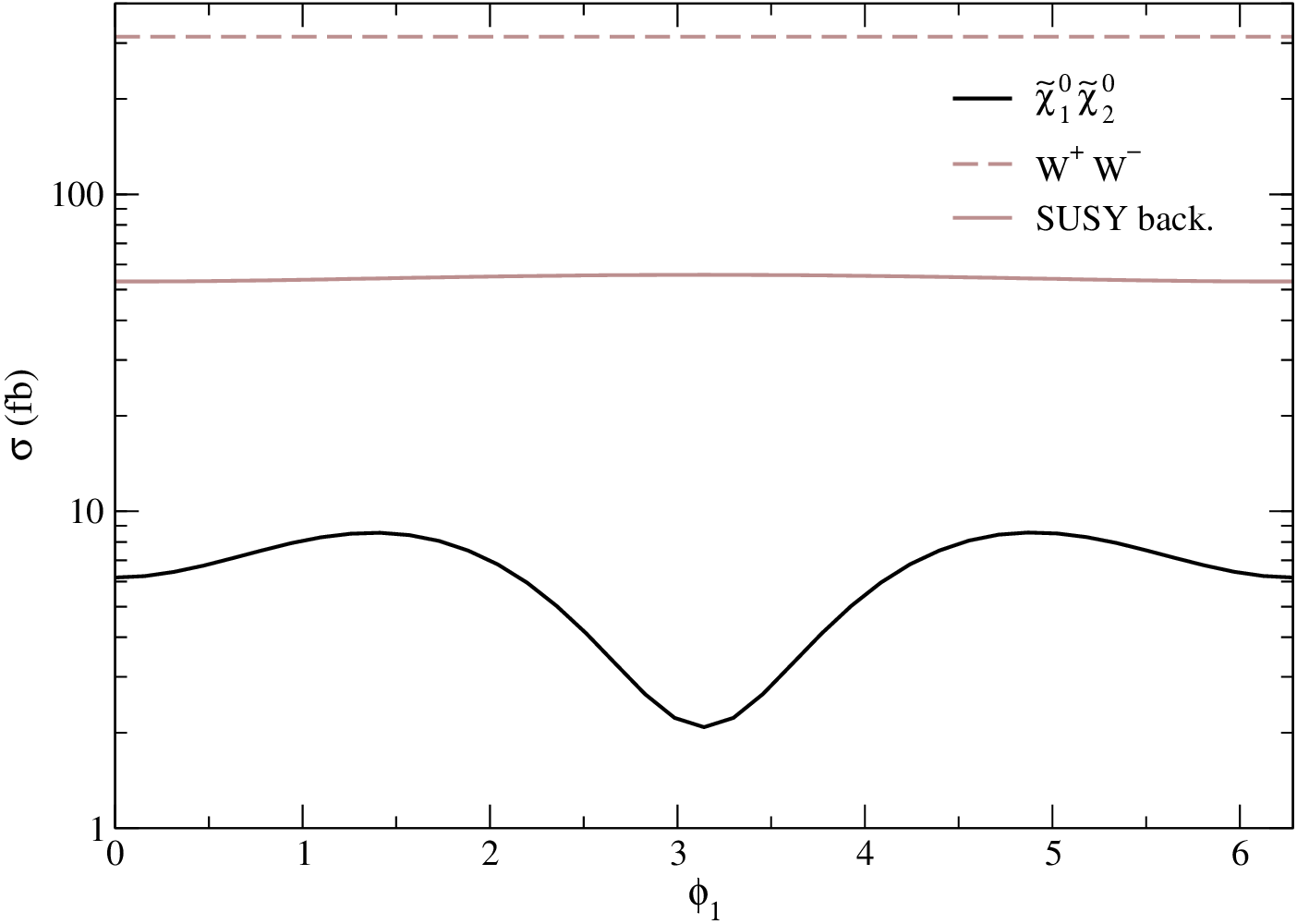,width=7.3cm,clip=} 
& \epsfig{file=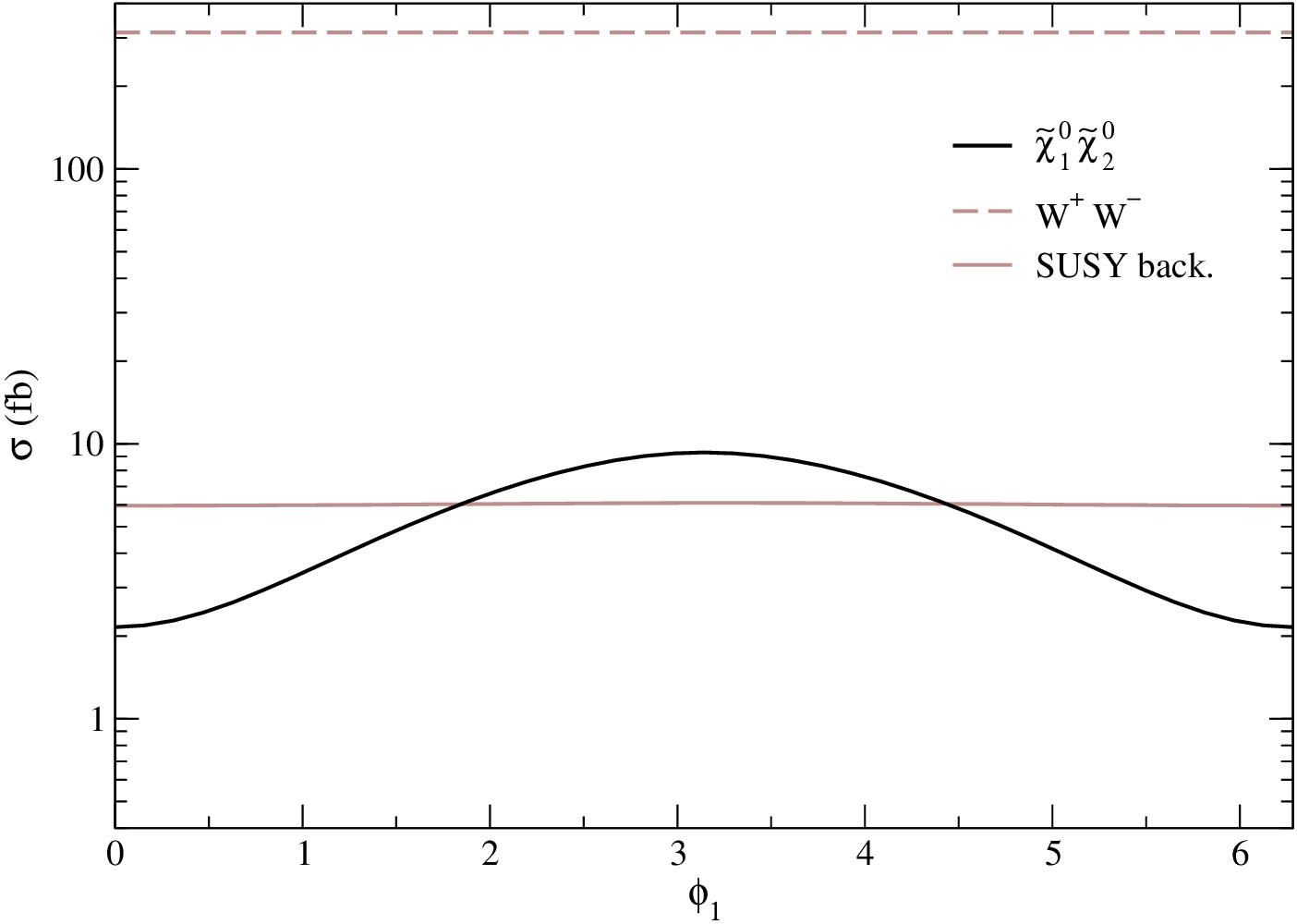,width=7.3cm,clip=} \\
(a) & (b)
\end{tabular}
\caption{Dependence of the signal and background cross sections on the phase
$\phi_1$ in scenario 1 (a) and scenario 2 (b), for $P_{e^+}=0.6$,
$P_{e^-}=-0.8$.}
\label{fig:crpm}
\end{center}
\end{figure}

\begin{figure}[htb]
\begin{center}
\begin{tabular}{cc}
\epsfig{file=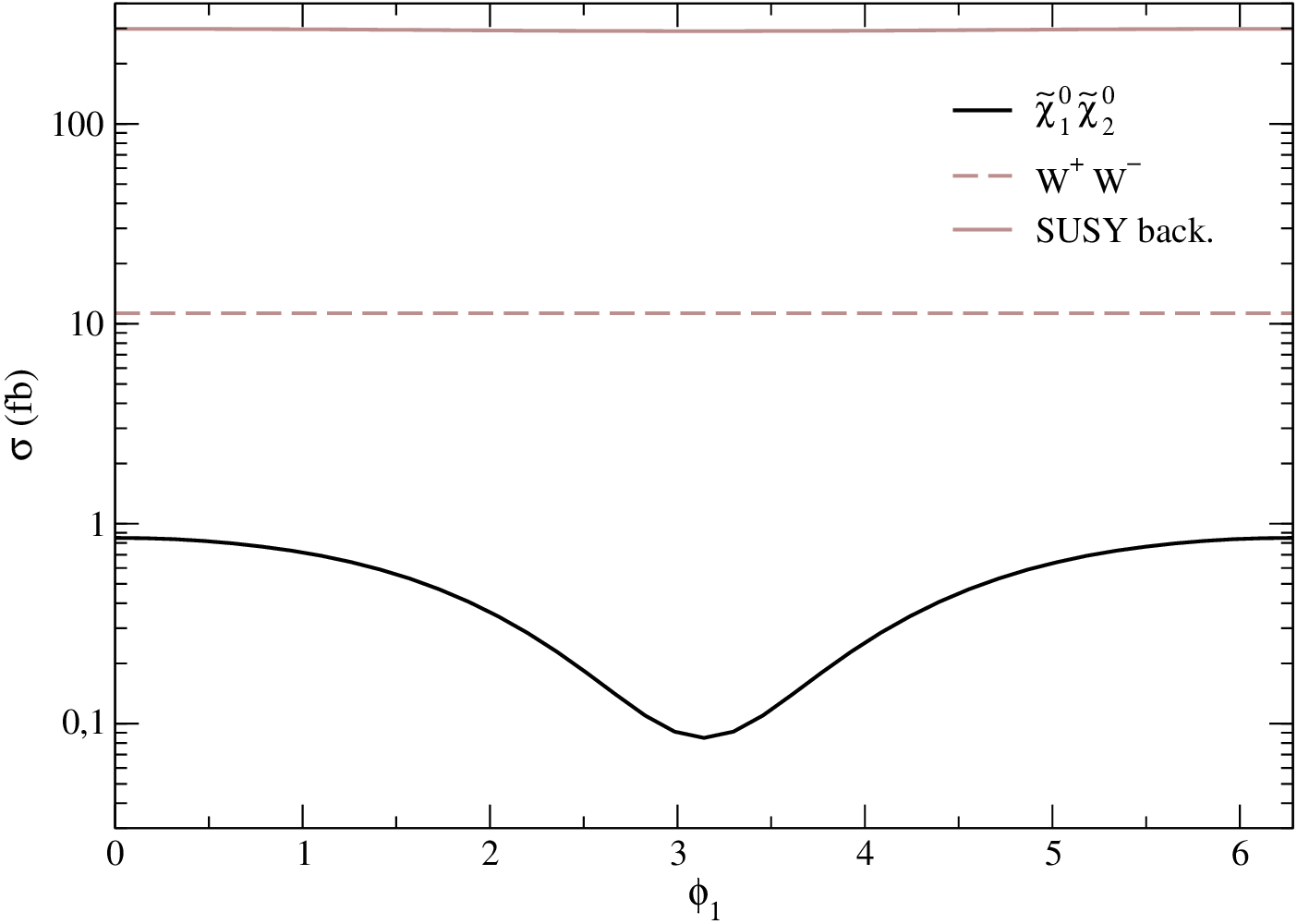,width=7.3cm,clip=} 
& \epsfig{file=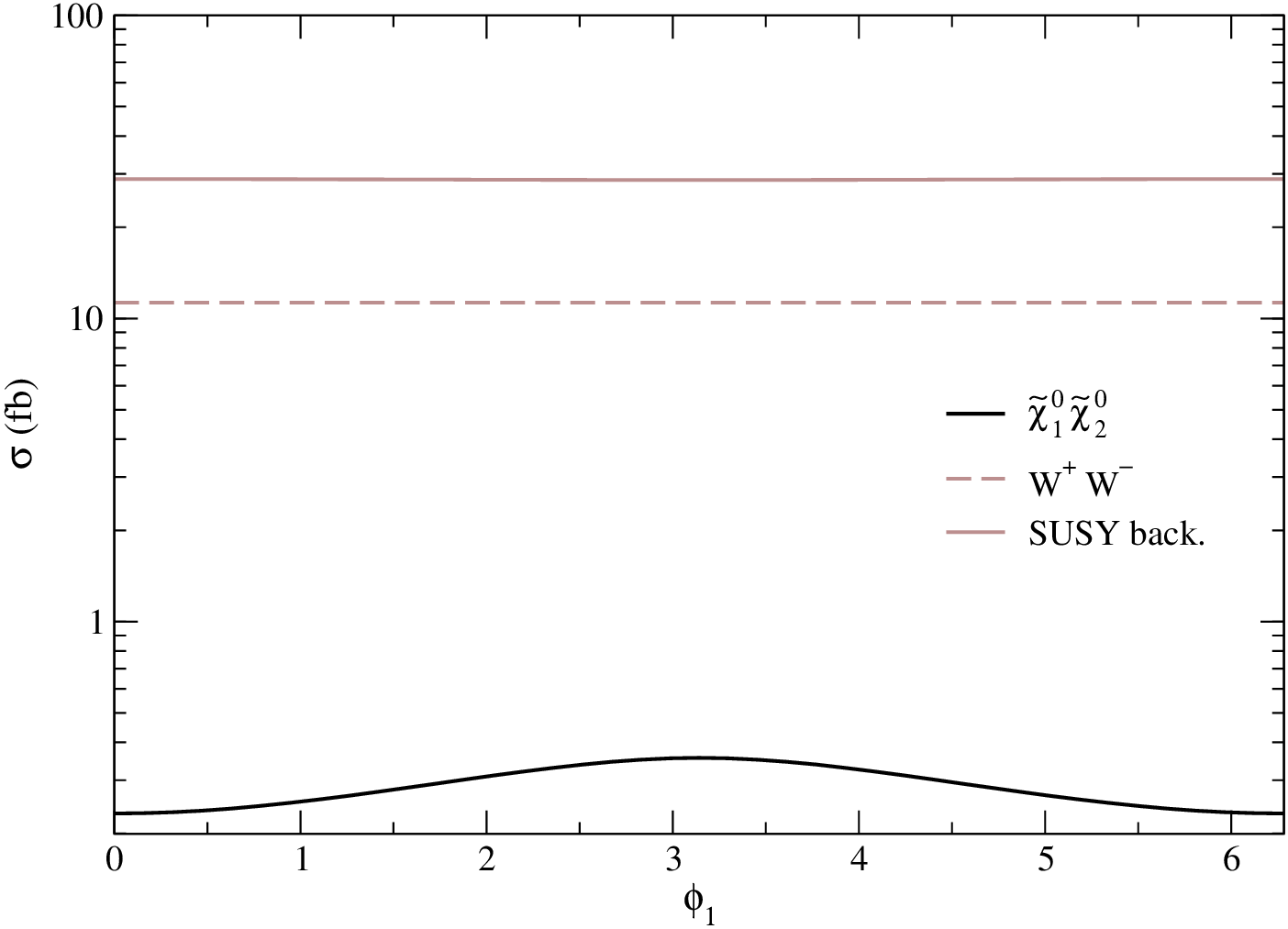,width=7.3cm,clip=} \\
(a) & (b)
\end{tabular}
\caption{Dependence of the signal and background cross sections on the phase
$\phi_1$ in scenario 1 (a) and scenario 2 (b), for $P_{e^+}=-0.6$,
$P_{e^-}=0.8$.}
\label{fig:crmp}
\end{center}
\end{figure}

The dependence of the asymmetry $A_1$ on the phase $\phi_1$ (taking $\phi_\mu =
0$) is shown in Fig.~\ref{fig:aXX} for the two scenarios under study
and three polarisation choices. In these plots we consider ISR,
beamstrahlung, beam spread and detector effects, but do not include backgrounds.
We observe that for $P_{e^+}=0.6$, $P_{e^-}=-0.8$ the asymmetry is
slightly larger than in the unpolarised case, while it is significantly reduced
for $P_{e^+}=-0.6$, $P_{e^-}=0.8$. For completeness, we show in
Fig.~\ref{fig:aXX-mu} the dependence of $A_1$ on $\phi_\mu$ for $\phi_1 = 0$.
For the two polarisation choices which turn out to be of interest for this
process ($P_{00}$ and $P_{+-}$) the asymmetry is virtually independent of
$\phi_\mu$. The cross sections of the signal and
backgrounds are plotted in Figs.~\ref{fig:cr00}--\ref{fig:crmp} as a function of
$\phi_1$. $\tilde
\chi_1^0 \tilde \chi_2^0$ production is
largest for $P_{e^+}=0.6$, $P_{e^-}=-0.8$ because the dominant contributions to
the amplitude are from $\tilde e_L$ exchange diagrams in both scenarios.
These polarisations enhance the $W^+ W^-$ cross section, which is the
largest background, but reduce $\tilde \ell_R \tilde \ell_R$ production which
is the second one in importance. It is clear from
Figs.~\ref{fig:aXX}--\ref{fig:crmp} that in both scenarios this polarisation
choice yields the best sensitivity for the measurement of $A_1$. We note here
that the determination of $\phi_1$ from a cross section measurement does not
seem possible, due not only to the small relative variation of the total
(signal plus background) cross section but also to the theoretical uncertainties
regarding neutralino mixing, sparticle mass spectrum, scale dependence of the
cross sections, etc.

\begin{figure}[htb]
\begin{center}
\begin{tabular}{cc}
\epsfig{file=Figs/asimXX-mu-2R.eps,width=7.3cm,clip=} 
& \epsfig{file=Figs/asimXX-mu-3.eps,width=7.3cm,clip=} \\
(a) & (b)
\end{tabular}
\caption{Dependence of the asymmetry $A_1$ on the phase $\phi_\mu$ in scenario 1
(a) and scenario 2 (b). Backgrounds are not included. We consider unpolarised
beams ($P_{00}$), $P_{e^+}=0.6$, $P_{e^-}=-0.8$ ($P_{+-}$) and
$P_{e^+}=-0.6$, $P_{e^-}=0.8$ ($P_{-+}$).}
\label{fig:aXX-mu}
\end{center}
\end{figure}

\begin{figure}[tb]
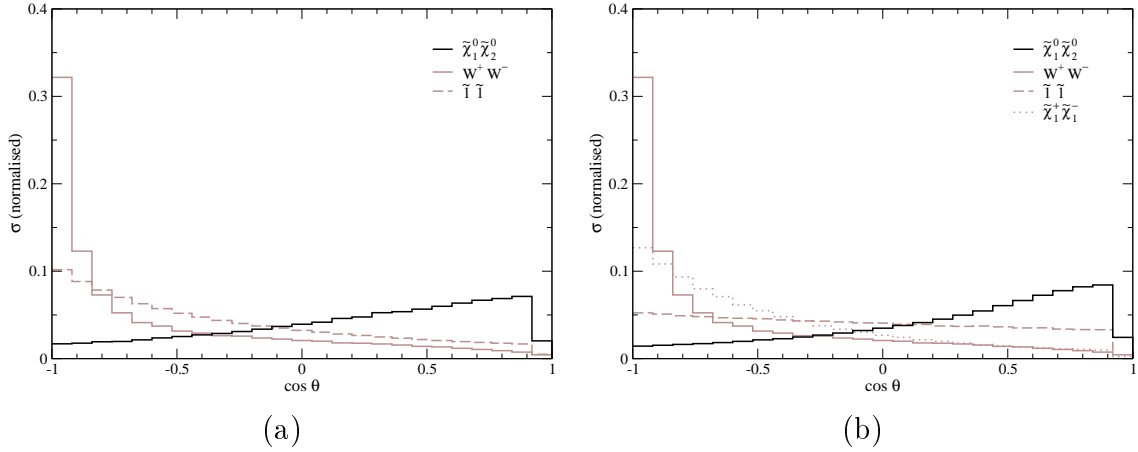

\begin{center}
\begin{tabular}{cc}
\epsfig{file=Figs/costh-2R.eps,width=7.3cm,clip=} 
& \epsfig{file=Figs/costh-3.eps,width=7.3cm,clip=} \\
(a) & (b)
\end{tabular}
\caption{Dependence of the signal and background cross sections on the angle
$\theta$ between $\ell^+$ and $\ell^-$ in
scenario 1 (a) and scenario 2 (b). We have set $\phi_1 = 0$ and
$P_{e^+}=0.6$, $P_{e^-}=-0.8$. All cross sections are normalised to unity.}
\label{fig:th}
\end{center}
\end{figure}

The $W^+ W^-$ background can be effectively reduced requiring that the angle
$\theta$ between the two final state charged leptons is smaller than, for
instance,
$90^\circ$ (we have not attempted to optimise the signal to background ratio
but rather we have chosen $\theta \leq 90^\circ$ in all cases for simplicity).
The kinematical distribution of the signal and backgrounds
with respect to $\cos \theta$ is shown in Fig.~\ref{fig:th}, with all cross
sections normalised to unity. The total cross sections are: $\sigma_{W^+ W^-} =
318$ fb; $\sigma_{\tilde \chi_1^0 \tilde \chi_2^0} =6.2 ~ (2.2)$ fb,
$\sigma_{\tilde \ell \tilde \ell} = 52.1 ~ (3.0)$ fb,
$\sigma_{\tilde \chi_1^+ \tilde \chi_1^-} = 1.2 \times 10^{-3} ~ (2.9)$ fb
in scenario 1 (scenario 2).
$W^+ W^-$ production is strongly peaked at $\theta = 180^\circ$, because $W^+
W^-$ pairs are produced with high momentum in the CM frame and
in the $W$ rest frame the charged lepton is preferrably emitted in the
direction of the $W$ boson CM momentum.
Slepton decays are isotropic, thus the only dependence on the angle $\theta$ of
the $\tilde \ell \tilde \ell$ cross section is kinematical. It can be noticed
that the decrease with $\cos \theta$ of their cross section is more pronounced
in scenario 1 (in this case the sleptons
are lighter and then produced with larger energy and momentum).

\begin{figure}[htb]
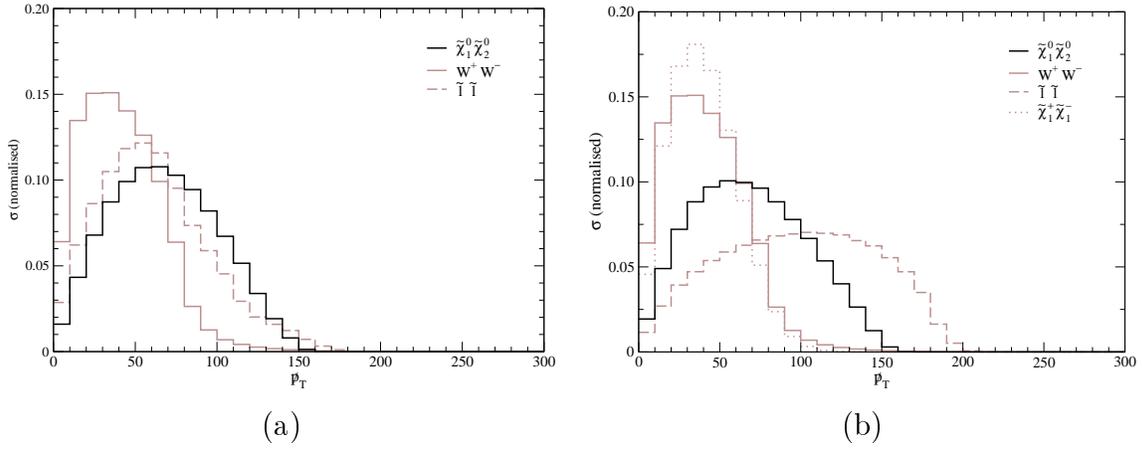

\begin{center}
\begin{tabular}{cc}
\epsfig{file=Figs/ptmiss-2R.eps,width=7.3cm,clip=} 
& \epsfig{file=Figs/ptmiss-3.eps,width=7.3cm,clip=} \\
(a) & (b)
\end{tabular}
\caption{Kinematical distribution of the missing transverse momentum $\ptmiss$
in scenario 1 (a) and scenario 2 (b). We have set $\phi_1=0$ and $P_{e^+}=0.6$,
$P_{e^-}=-0.8$. All cross sections are normalised to unity.}
\label{fig:ptmiss}
\end{center}
\end{figure}

\begin{figure}[htb]
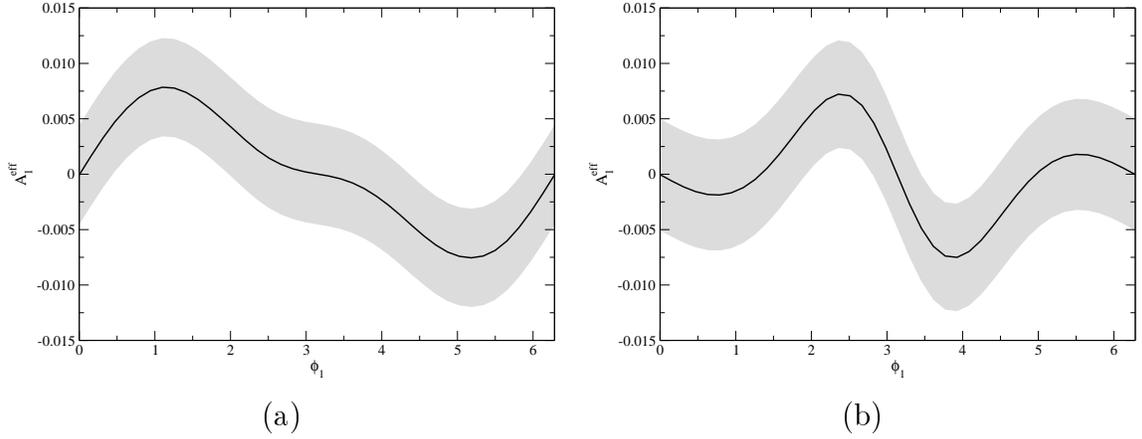

\begin{center}
\begin{tabular}{cc}
\epsfig{file=Figs/asim-2R.eps,width=7.3cm,clip=} &
\epsfig{file=Figs/asim-3.eps,width=7.3cm,clip=} \\
(a) & (b)
\end{tabular}
\caption{Dependence of the observed asymmetry $A_1^\mathrm{eff}$ (including
backgrounds) on the phase $\phi_1$ in scenario 1 (a) and scenario 2 (b), for
$P_{e^+}=0.6$ and $P_{e^-}=-0.8$. The shaded area represents the statistical
error for two years of running.}
\label{fig:asim}
\end{center}
\end{figure}

One could expect that the $W^+ W^-$ background had smaller values of the missing
transverse momentum $\ptmiss$ than SUSY processes, in which there is a pair of
heavy undetected $\tilde \chi_1^0$ in the final state. However, as we observe in
Fig.~\ref{fig:ptmiss}, the kinematical distributions are not so different, and
trying to reduce the background requiring large $\ptmiss$ eliminates a large
fraction of the signal. The results for the observed asymmetry
$A_1^\mathrm{eff}$ (including backgrounds) are presented in Fig.~\ref{fig:asim}
for both scenarios, using $P_{e^+}=0.6$, $P_{e^-}=-0.8$, which give the
best results, and requiring $\theta \leq 90^\circ$. We also show the statistical
error for two years with a luminosity of 345 fb$^{-1}$ per year. Comparing with
Fig.~\ref{fig:aXX} we see that the asymmetry roughly decreases by a factor
$r \sim 1/10$ due to the backgrounds, but it can be observed after a few
years of running for wide ranges of $\phi_1$. We remark that detector and
kinematical cuts do not generate by themselves a ``fake'' asymmetry $A_1$ (the
asymmetry goes to zero when CP is conserved). Still, the cut $\theta \leq
90^\circ$ slightly enhances the values of $A_1$ when they are nonzero.

We collect in Table~\ref{tab:cross} the signal and background cross sections for
two examples in which the observability of the CP asymmetry is nearly maximal:
$\phi_1 = \pi/3$ in scenario 1 and $\phi_1 = 3 \pi/4$ in scenario 2. The
kinematical cut on $\theta$ reduces $W^+ W^-$ and $\tilde \ell \tilde \ell$
production by factors of 6 and $2-3$, respectively, while keeping approximately
70\% of the signal. The CP asymmetries before cuts in these scenarios
are $A_1 = 0.101$, $A_1 =
0.066$, and after cuts they are $A_1 = 0.108$, $A_1 = 0.074$. The ratio $r$
after cuts is $r = 0.074$, $r = 0.097$, yielding effective asymmetries
$A_1^\mathrm{eff} = 0.0080$, $A_1^\mathrm{eff} = 0.0072$.
These asymmetries can be observed with statistical significances of $1.8 \,
\sigma$ and $1.5 \, \sigma$, respectively, after two years of running.

\begin{table}[htb]
\begin{center}
\begin{tabular}{ccccc}
& \multicolumn{2}{c}{Scenario 1} & \multicolumn{2}{c}{Scenario 2} \\
& before & after & before & after \\
\hline
$\tilde \chi_1^0 \tilde \chi_2^0$ & 8.18 & 5.49 & 7.83 & 6.01 \\
$W^+ W^-$ & 318.0 & 54.3 & 318.0 & 54.3 \\
$\tilde \ell^+ \tilde \ell^-$ & 53.1 & 14.2 & 3.11 & 1.33 \\
$\tilde \chi_1^+ \tilde \chi_1^-$ & \multicolumn{2}{c}{$\sim 10^{-3}$}
  & 2.96 & 0.56
\end{tabular}
\caption{Cross sections (in fb) before and after kinematical cuts of the signal
and backgrounds in scenario 1 (a) and scenario 2 (b). In the first scenario we
choose $\phi_1 = \pi/3$, and in the second $\phi_1 = 3 \pi/4$.}
\label{tab:cross}
\end{center}
\end{table}

\subsection{Other SUSY scenarios}

In the two scenarios analysed in detail the asymmetry $A_1$ is difficult to
observe due to the fact that the same beam polarisations 
$P_{e^+}=0.6$, $P_{e^-}=-0.8$ which make the signal largest also enhance the
most important background, which is $W^+ W^-$ production, and for 
$P_{e^+}=-0.6$, $P_{e^-}=0.8$ the signal is small, because $\tilde e_L$
exchange dominates the amplitudes of $e^+ e^- \to \tilde \chi_1^0 \tilde
\chi_2^0$. One can then
wonder what is the situation in SUSY scenarios where $\tilde e_R$ exchange is
important, so that $\tilde \chi_1^0 \tilde \chi_2^0$ production is large
with the latter choice of beam polarisations.
The situation does not improve however, because these polarisations increase the
cross section for $\tilde \ell_R \tilde \ell_R$ production, which is the second
largest background and more difficult to remove with kinematical cuts
(see Figs.~\ref{fig:th},\ref{fig:ptmiss}). We have explicitly analysed one
scenario with large neutralino mixing, finding results very similar
to those presented in the previous subsection. The parameters for this scenario
are: $M_1 = 105.8 \, e^{i \phi_1}$ GeV, $M_2 = 199.3$ GeV, $\mu = 263.5 \,
e^{i \phi_\mu}$ GeV, $m_{\tilde L} = m_{\tilde E} = 100$ GeV,
$A_E = 0$, $\tan \beta = 10$. The relevant sparticle
masses are $m_{\tilde e_R,\tilde \mu_R} = 147.5$ GeV,
$m_{\tilde e_L,\tilde \mu_L} = 211.5$ GeV,
$m_{\tilde \chi_1^0} \simeq 103$ GeV, $m_{\tilde \chi_2^0} \simeq 174$ GeV,
close to the values for scenario 1. We have found that in the best case the
asymmetry can be observed with $1.2 \, \sigma$ after two years of running. 
More favourable SUSY scenarios may be found, but the general
trend is that the CP asymmetry $A_1$ is difficult to observe, due either to the
$W^+ W^-$ background or the $\tilde \ell_R \tilde \ell_R$ background.

\section{Summary and conclusions}
\label{sec:4}

The determination of the presence (or not) of complex phases in the neutralino
sector is one of the tasks that must be carried out at a future linear collider.
This will be done following two different approaches: with a precise analysis
of CP-conserving quantities (see e.g, Refs.~\cite{CPcon1,CPcon2,choi}) and
through the investigation
of CP-violating asymmetries. We have shown that in $\tilde \chi_1^0 \tilde
\chi_2^0$ production it is possible to have a CP asymmetry in the triple product
$\vec p_{e^+} \cdot ( \vec p_{\ell^-} \times \vec p_{\ell^+} )$
which is very sensitive to the phase $\phi_1$ of the gaugino mass $M_1$.
We have studied two SUSY scenarios, one with $\tilde \chi_2^0 \to \tilde
\ell_R^\pm \ell^\mp$ and the other with three-body decays of $\tilde \chi_2^0$.
In both scenarios the neutralinos are light enough to be accessible at a
CM energy of 500 GeV, as proposed for the first phase of TESLA and NLC. 
The CP asymmetries in $\tilde \chi_1^0 \tilde \chi_2^0$ production are of order
0.1, and the ``effective'' asymmetries observed (which include the backgrounds)
are roughly one order of magnitude smaller. At any rate, asymmetries of order
0.01 can be observed after a few years of running with the planned luminosity,
for wide intervals of $\phi_1$.
The results for a heavier SUSY spectrum are similar, and could be experimentally
studied with higher CM energies and luminosities.

It should be emphasised that this study is complementary to
the analysis of CP asymmetries in other processes.
Selectron cascade decays $\tilde e_L \to e \tilde \chi_2^0 \to e \tilde
\chi_1^0 \mu^+ \mu^-$  are one example, where there may exist a CP
asymmetry in the triple product
$\vec s \cdot (\vec p_{\mu^-} \times \vec p_{\mu^+})$, with $\vec s$ 
the $\tilde \chi_2^0$ spin \cite{casc}. In the scenario with $\tilde \chi_2^0$
three-body
decays discussed, it is easier to observe a CP asymmetry in the latter process.
In particular, the asymmetry in $\tilde \chi_1^0 \tilde \chi_2^0$ production
is negligible for $\phi_1=\pi/2$, $\phi_1 = 3 \pi/2$, whereas in selectron
decays it is nearly maximal. The reverse situation occurs in the scenario with
decays $\tilde \chi_2^0 \to \tilde \ell_R^\pm \ell^\mp$: the asymmetry in
selectron decays is very small, while it could be observable in
$\tilde \chi_1^0 \tilde \chi_2^0$ production for values of $\phi_1$ around
$\pm 1$. The combined analysis of these and other processes, together with the
constraints from EDMs, may allow the determination of the CP-violating phases
in the neutralino sector.

\vspace{1cm}
\noindent
{\Large \bf Acknowledgements}

\vspace{0.4cm} \noindent
This work has been supported
by the European Community's Human Potential Programme under contract
HTRN--CT--2000--00149 Physics at Colliders and by FCT
through projects POCTI/FNU/43793/2002, CFIF--Plurianual (2/91) and
grant SFRH/ BPD/12603/2003.

\appendix
\section{Notation and conventions}
\label{sec:A}
\setcounter{equation}{0}
\renewcommand{\theequation}{\thesection.\arabic{equation}}

We list here some of the mass matrices and interactions used in this work,
(see also Ref.~\cite{npb}) following the conventions of Ref.~\cite{romao}.
We neglect flavour mixing and assume that the trilinear terms are real.

The relation between slepton
mass eigenstates $\tilde \ell = ( \tilde \ell_1 ~ \tilde \ell_2 )^T$
(with $m_{\tilde \ell_1} < m_{\tilde \ell_2}$)
and weak interaction eigenstates $\tilde \ell' = ( \tilde \ell_L ~ \tilde
\ell_R )^T$
can be written as $\tilde \ell = R^{\tilde \ell} \, \tilde \ell'$, with
\begin{equation}
R^{\tilde \ell} =
 \left( \begin{array}{cc} \sin\theta_{\tilde \ell} & \cos \theta_{\tilde \ell}
  \\
 -\cos \theta_{\tilde \ell} & \sin \theta_{\tilde \ell} \end{array} \right) \,.
\end{equation}
In the basis where $\psi^+ = (\tilde W^+,\tilde H_2^+)^T$,
$\psi^- = (\tilde W^-,\tilde H_1^+)^T$, the chargino mass term is
\begin{equation}
\mathcal{L}_{\tilde \chi^\pm}^\mathrm{mass} = - (\psi^-)^T M_{\tilde \chi^\pm}
\, \psi^+ + \mathrm{H.c.} \,,
\end{equation}
being the chargino mass matrix
\begin{equation}
M_{\tilde \chi^\pm} = \left( \begin{tabular}{cc}
$M_2$ & $\sqrt 2 \, m_W \sin \beta$ \\ $\sqrt 2 \, m_W \cos \beta$ & $\mu$
\end{tabular} \right) \,.
\end{equation}
This matrix can be diagonalised with two unitary matrices $U$ and $V$,
\begin{equation}
U^* M_{\tilde \chi^\pm} V^\dagger = M_{\tilde \chi^\pm}^\mathrm{diag} \,.
\end{equation}
The physical chargino fields are $\tilde \chi_i^- = \left( \chi_i^- ~
\overline{\chi_i^+} \right)^T$,
with $\chi^- = U \psi^-$, $\chi^+ = V \psi^+$.
Their couplings to leptons are
\begin{eqnarray}
\mathcal{L}_{\tilde \nu_\ell \ell \tilde \chi_j^-} & = & \tilde \nu_\ell \; \bar
  \ell \left[ B_{Lj}^\ell P_L + B_{Rj}^\ell P_R \right] \tilde \chi_j^- 
  + \mathrm{H.c.} \,, \nonumber \\
\mathcal{L}_{\tilde \ell_i \nu_\ell \tilde \chi_j^-} & = & \tilde \ell_i^*
  \; \bar \nu_\ell \left[ B_{Lij}^\nu P_L \right] \tilde \chi_j^-
  + \mathrm{H.c.} \,,
\end{eqnarray}
with
\begin{eqnarray}
B_{Lj}^\ell & = & Y_\ell \, U_{j2}^* \,, \nonumber \\
B_{Rj}^\ell & = & -g \, V_{j1} \,, \nonumber \\
B_{Lij}^\nu & = & -g \, U_{j1}^* \, R_{i1}^{\tilde \ell} + Y_\ell \, U_{j2}^* \,
R_{i2}^{\tilde \ell}  \,.
\end{eqnarray}
For $\ell = e,\mu$ the terms with the Yukawa coupling $Y_\ell$ can be safely
neglected. The chargino interactions with the gauge bosons are
\begin{eqnarray}
\mathcal{L}_{\gamma \tilde \chi_i^- \tilde \chi_i^-} & = & e \, A_\mu \,
  \overline{\tilde \chi_i^-} \gamma^\mu \tilde \chi_i^- \,, \nonumber \\
\mathcal{L}_{Z \tilde \chi_i^- \tilde \chi_j^-} & = & \frac{g}{2 \cos \theta_W}
  \, Z_\mu \left[ \overline{\tilde \chi_i^-} \gamma^\mu \left( E_L^{ij} P_L +
   E_R^{ij} P_R \right) \tilde \chi_j^- \right] \,, \nonumber \\
\mathcal{L}_{W \tilde \chi_i^0 \tilde \chi_j^-} & = & g \, W_\mu^\dagger
  \left[ \overline{\tilde \chi_i^0} \gamma^\mu \left( O_L^{ij} P_L + O_R^{ij}
   P_R \right) \tilde \chi_j^- \right] + \mathrm{H.c.} \,,
\end{eqnarray}
with
\begin{eqnarray}
E_L^{ij} & = & U_{i2} \, U_{j2}^* + 2 U_{i1} \, U_{j1}^* - 2 \delta_{ij} \,
  \sin^2 \theta_W \,, \nonumber \\
E_R^{ij} & = & V_{j2} \, V_{i2}^* + 2 V_{j1} \, V_{i1}^* - 2 \delta_{ij} \,
  \sin^2 \theta_W \,, \nonumber \\
O_L^{ij} & = & -N_{i2} \, U_{j1}^* - \frac{1}{\sqrt 2} N_{i3} \, U_{j2}^* \,,
  \nonumber \\
O_R^{ij} & = & -N_{i2}^* \, V_{j1} + \frac{1}{\sqrt 2} N_{i4}^* \, V_{j2}
\end{eqnarray}
and $\theta_W$ the weak mixing angle.
Slepton couplings to the neutral gauge bosons are given by
\begin{eqnarray}
\mathcal{L}_{\gamma \tilde \ell_i \tilde \ell_i} & = & -i e \, A_\mu \, \left[
  \tilde \ell_i^* \, \overleftrightarrow{\partial_\mu}  \, \tilde \ell_i
  \right] \,, \nonumber \\
\mathcal{L}_{Z \tilde \ell_i \tilde \ell_i} & = & -i \frac{g}{2 \cos \theta_W}
  \, Z_\mu \left[
    z^\ell_{ij} \, \tilde \ell_i^* \, \overleftrightarrow{\partial_\mu} \,
      \tilde \ell_j
    \right] \,, \nonumber \\
\mathcal{L}_{Z \tilde \nu_\ell \tilde \nu_\ell} & = & -i \frac{g}{2 \cos
 \theta_W} \, Z_\mu \left[ \tilde \nu_\ell^* \,
  \overleftrightarrow{\partial_\mu}  \, \tilde \nu_\ell \right] \,.
\end{eqnarray}
The $z^{\ell}_{ij}$ mixing parameters read
\begin{eqnarray}
z^\ell_{11} & = & \left( -1+2 \sin^2 \theta_W \right) |R_{11}^{\tilde \ell}|^2
+ 2 \sin^2 \theta_W |R_{12}^{\tilde \ell}|^2 \,, \nonumber \\
z^\ell_{22} & = & \left( -1+2 \sin^2 \theta_W \right) |R_{21}^{\tilde \ell}|^2
+ 2 \sin^2 \theta_W |R_{22}^{\tilde \ell}|^2 \,, \nonumber \\
z^\ell_{12} & = & - R^{\tilde \ell}_{11} R^{\tilde \ell *}_{21} \,, \nonumber \\
z^\ell_{21} & = & z^{\ell*}_{12} \,.
\end{eqnarray}
Finally, the neutrino--sneutrino--neutralino couplings are
\begin{equation}
\mathcal{L}_{\tilde \nu_\ell \nu_\ell \tilde{\chi}_j^0} = \tilde \nu_\ell^* \,
  \overline{\tilde \chi_j^0} \left[ C^\nu_{Lij} \, P_L \right] \nu_\ell +
  \tilde \nu_\ell \, \bar \nu_\ell \left[ C^{\nu*}_{Lij} \, P_R \right]
  \tilde \chi_j^0 \,,
\end{equation}
where
\begin{equation}
C^\nu_{Lij} = -\frac{g}{\sqrt 2} \left( N^*_{j2} - \tan \theta_W N^*_{j1}
\right) \,.
\end{equation}

\end{document}